\documentclass[11pt]{article}
\usepackage{graphicx}
\usepackage{amsmath}
\usepackage{amssymb}
\usepackage{caption2}
\begin{document}
\bibliographystyle{prsty}
\begin{center}
{\large {\bf \sc{  Analysis of the decays $\psi^{\prime}\to
{J/\psi}\pi^+\pi^-$ and $\eta_c^{\prime}\to {\eta_c}\pi^+\pi^-$ with the heavy quark symmetry
 }}} \\[2mm]
Zhi-Gang Wang \footnote{E-mail,wangzgyiti@yahoo.com.cn.  }    \\
 Department of Physics, North China Electric Power University, Baoding 071003, P. R. China
\end{center}

\begin{abstract}
In this article, we study the decays $\psi^{\prime}\to
{J/\psi}\pi^+\pi^-$ and $\eta_c^{\prime}\to {\eta_c}\pi^+\pi^-$   by
taking into account  the chiral symmetry breaking effects,  the
final-state interactions and the heavy quark symmetry.  We can
confront  the predictions of the $\eta_c^{\prime}\to \eta_c
\pi^+\pi^-$ decay width and differential decay width with  the
experimental data in the future, and obtain  powerful constraints on
the chiral breaking effects and the final-state interactions, and
test the heavy quark symmetry.
\end{abstract}



\section{Introduction}

Hadronic transitions among the charmonium and bottomonium states
$\psi({\rm{mS}})\to\psi({\rm{nS}})\pi^+\pi^-$ and
$\Upsilon({\rm{mS}})\to\Upsilon({\rm{nS}})\pi^+\pi^-$  are of
particular  interesting for studying the dynamics of both the heavy
quarkonia and the light mesons.  Such processes are usually
calculated with the multipole expansion in QCD, where the heavy
quarkonia are considered as the compact and nonrelativistic objects
and emit soft gluons which hadronize into the light meson or light
meson pair \cite{QCDPE}. The amplitudes  can be factorized into the
heavy quarkonium part  and the light meson part. The former part
depends on the dynamics of the quarkonium and should preserve the
heavy quark symmetry, while the latter part depends on the chiral
dynamics and should obey the chiral symmetry
\cite{Chiral-H,Voloshin-2pi,Charmonium-Review}.

In Ref.\cite{Mannel}, Mannel and Urech  construct an effective
Lagrangian for the hadronic decays $\psi^{\prime}\to
{J/\psi}\pi^+\pi^-$ and $\Upsilon^{\prime}\to {\Upsilon}\pi^+\pi^-$
based on the heavy quark symmetry and the chiral symmetry, and
obtain reasonable values for the coupling constants by fitting to
the invariant $\pi^+\pi^-$ mass distributions. In
Ref.\cite{ZhuangTL}, Yan, Wei and Zhuang observe  that there are
$D$-wave contributions besides the $S$-wave contributions, although
the $D$-wave contributions are very small. The $D$-wave
contributions  were  firstly predicted  by the Novikov-Shifman model
based on the multipole expansion in QCD combined with  the chiral
symmetry, current algebra and partially conserved axial-vector
current \cite{D-wave}. In Refs.\cite{FKGuo2-1,FKGuo2-2}, Guo et al
observe that the final-state interactions play an important role. In
the case of the transitions
$\Upsilon({\rm{3S}})\to\Upsilon({\rm{1S}})\pi\pi$, there is a double
peak in the $\pi\pi$ invariant mass spectrum \cite{CLEO-Y}, we have
to resort to additional assumptions to describe the experimental
data, such as the relativistic corrections \cite{Voloshin-2pi}, the
final-state interactions and the $f_0(600)$ resonance
\cite{FSI-Y-1,FSI-Y-2}, the exotic $\Upsilon \pi$ resonances
\cite{FKGuo2-1,FSI-Y-1,exotic-Y}, the coupled channel effects
\cite{Couple-Y}, the $\rm{S}-\rm{D}$ mixing \cite{SD-Y}, the field
correlators \cite{FCM-Y},  etc.

The two $\pi$ transition  $\eta_c^{\prime}\to {\eta_c}\pi^+\pi^-$
has not been observed yet. Recently, the CLEO collaboration searched
for the decay $\psi^{\prime} \to \gamma \eta_c^{\prime}$ in a sample
of 25.9 million $\psi^{\prime}$ events collected with the CLEO-c
detector, and observed no evidence for the decays $\psi^{\prime} \to
\gamma \eta_c^{\prime}$ and  $\eta_c^{\prime}\to
{\eta_c}\pi^+\pi^-$, and set  the upper limit,
\begin{eqnarray}
{\rm{Br}}(\psi^{\prime}\to\gamma
\eta_c^{\prime})\times{\rm{Br}}(\eta_c^{\prime}\to\eta_c\pi^+\pi^-)<1.7\times
10^{-4}\, ,
\end{eqnarray}
at the $90\%$ confidence level \cite{CLEO-No-etac}. It is
interesting to make predictions for the decay width and differential
decay width of the process  $\eta_c^{\prime}\to {\eta_c}\pi^+\pi^-$,
which may be observed at the BESIII and $\bar{{\rm P}}\rm{ANDA}$ in
the future  \cite{BESIII,PANDA}.

In this article, we study the decays $\psi^{\prime}\to
{J/\psi}\pi^+\pi^-$ and $\eta_c^{\prime}\to {\eta_c}\pi^+\pi^-$ with
a phenomenological Lagrangian  by taking into account  the chiral
symmetry breaking effects, the final-state interactions and the
heavy quark symmetry
\cite{Mannel,ZhuangTL,FKGuo2-1,FKGuo2-2,GattoPLB93,PRT1997}.

The article is arranged as follows:  we study the $\pi^+\pi^-$
transitions of the $\psi^{\prime}$ and $\eta_c^{\prime}$ in details
based on the heavy quark symmetry in Sec.2; in Sec.3, we present the
numerical results and discussions; and Sec.4 is reserved for our
conclusions.

\section{ The $\pi^+\pi^-$ transitions of the  $\psi^{\prime}$ and $\eta_c^{\prime}$ with the heavy quark symmetry }

The  charmonium states can be classified according to the
 notation ${\rm n}^{2s+1}L_{j}$, where the ${\rm n}$ is the radial quantum number, the $L$ is
  the orbital angular momentum, the $s$ is  the spin, and the $j$ is the total angular
momentum. They have the parity and charge conjugation $P=(-1)^{L+1}$
and $C=(-1)^{L+s}$, respectively. The states have the same radial
quantum number ${\rm n}$ and  orbital momentum $L=0$ can be
expressed by  the superfield $J$ \cite{GattoPLB93,PRT1997},
\begin{eqnarray}
J&=&\frac{1+{\rlap{v}/}}{2}\left[\psi_{\mu}\gamma^\mu-\eta_c\gamma_5\right]
\frac{1-{\rlap{v}/}}{2} \, ,
\end{eqnarray}
where the $v^{\mu}$ denotes the four velocity associated to the
superfield. We multiply the charmonium fields $\psi_{\mu}$ and
 $\eta_{c}$ with a factor $\sqrt{M_{\psi}}$  and $\sqrt{M_{\eta_c}}$
 respectively,  and they have dimension of mass
$\frac{3}{2}$.  The superfields have been used to construct the
phenomenological Lagrangians to study the radiative transitions,
pseudoscalar meson transitions and vector meson transitions among
the heavy quarkonia \cite{GattoPLB93,PRT1997,HH-transition}.

The $\pi^+\pi^-$ transitions between the ${\rm m}$ and ${\rm n}$
charmonium states  can be described by the following
phenomenological Lagrangian \cite{Mannel,PRT1997},
\begin{eqnarray}
{\cal{L}}&=&\frac{1}{2}\sum_{m,n}
\mathrm{Tr}\left[\bar{J}(m)J(n)\right]
\left\{g_1(m,n)\mathrm{Tr}[(\partial_{\alpha}U)(\partial^{\alpha}U)^\dag]
+ g_2(m,n)\mathrm{Tr}[(v\cdot{\partial}U)(v\cdot{\partial}U)^\dag]\right.\nonumber \\
&&\left.+ g_3(m,n)\mathrm{Tr}[M(U+U^{\dag}-2)] \right\} \, ,
\end{eqnarray}
where $\bar{J}=\gamma^0 J^{\dag} \gamma^0$, $M=B_0\,{\rm
{diag}}\{m_u, m_d, m_s\}$ with $m_\pi^2=2B_0m$, $m_K^2=B_0(m+m_s)$,
$m_\eta^2=\frac{2}{3}B_0(m+2m_s)$ in the isospin symmetry limit
$m_u=m_d=m$,  the $U$ is a $3{\times}3$ matrix that contains the
pseudoscalar Goldstone fields, and the $g_i(m,n)$ denote the
coupling constants. We construct the chiral symmetry breaking term
$\frac{1}{2}g_3(m,n)
\mathrm{Tr}\left[\bar{J}(m)J(n)\right]\mathrm{Tr}[M(U+U^{\dag}-2)]$
consulting Ref.\cite{Mannel}. In the following, we will smear the
indexes  $(m,n)$ in the coupling constants $g_i(m,n)$ for
simplicity.

The tree diagram amplitudes for the $\pi^+\pi^-$ transitions of the
$\psi^{\prime}$ and $\eta_c^{\prime}$ can be written  as
\begin{eqnarray}
T^0_{\psi'\to J/\psi \pi\pi}&=&-\frac{4}{f^{2}_\pi}\left[g_1p_1\cdot
p_2+g_{2}p_1^0p_2^0+g_3m_\pi^2\right]\sqrt{M_{\psi'}M_{J/\psi}}\epsilon^*\cdot\epsilon^{\prime} \, ,\nonumber\\
T^0_{\eta_c'\to
\eta_c\pi\pi}&=&+\frac{4}{f^{2}_\pi}\left[g_1p_1\cdot
p_2+g_{2}p_1^0p_2^0+g_3m_\pi^2\right]\sqrt{M_{\eta_c'}M_{\eta_c}}\,,
\end{eqnarray}
where the $\pi$ decay constant $f_\pi=92\,\rm{MeV}$, the $p_1$ and
$p_2$ are the four-momenta of the $\pi^+$ and $\pi^-$ respectively,
the $p_1^0$ and $p_2^0$ are the energies of the $\pi^+$ and $\pi^-$
in the laboratory frame respectively, and the $\epsilon $ and
$\epsilon^{\prime}$ are the polarization vectors of the
$\psi^{\prime}$ and $J/\psi$ respectively.
 The $p_1^0$ and
$p_2^0$ can be written as functions of the momenta of pions in the
center of mass frame of the $\pi\pi$ system:
\begin{eqnarray}
p_1^0 &=& \frac{1}{\sqrt{1-{ \beta}^2}}(E_1^{*}+|{
\mathbf \beta}||{\mathbf p}_1^{*}|\cos\theta_{\pi}^*)\, , \nonumber \\
p_2^0 &=& \frac{1}{\sqrt{1-{ \beta}^2}}(E_1^{*}-|{ \mathbf \beta
}||{\mathbf p}_1^{*}|\cos\theta_{\pi}^*) \, ,
\end{eqnarray}
where the ${\mathbf{\beta}}$ is the velocity of the $\pi\pi$ system
in the center of mass frame of the initial particle, the
$p_1^*$=($E_1^{*}$, ${\mathbf p}_1^*$) and $p_2^* $=($E_2^{*}$,
${\mathbf p}_2^*$) are the four-momenta of the $\pi^+$ and $\pi^-$
in the center of mass frame of the $\pi\pi$ system respectively. The
$p_1^0p_2^0$ can be written  as
\begin{eqnarray}
p_1^0p_2^0=\frac{1}{1-{ \mathbf \beta}^2}\left[\left(E_1^{*2}-\frac{{  \beta}^2{\mathbf p}_1^{*2}%
}{3}\right)P_0(\cos\theta_{\pi}^*)-2{ \mathbf \beta}^2{\mathbf
p}_1^{*2}P_2(\cos\theta_{\pi}^*)\right] \, ,
\end{eqnarray}
where $|{\bf p_1^*}|=|{\bf
p_2^*}|=\sqrt{\frac{m_{\pi\pi}^2}{4}-m_{\pi}^2}$, the
$P_0(\cos\theta_{\pi}^*)=1$ and
$P_2(\cos\theta_{\pi}^*)=\frac{1}{2}(\cos^2\theta_{\pi}^*-\frac{1}{3})$
are the Legendre functions \cite{FKGuo2-1,FKGuo2-2}.

In Refs.\cite{FKGuo2-1,FKGuo2-2}, Guo et al retain  the   coupling
constants $g_1$ and $g_2$, and  observe that the $S$-wave $\pi \pi$
final-state interactions  play an important role and should be
properly included. The chiral unitary theory is a suitable approach
for taking into account  the infinite series of the  re-scattering
meson loops \cite{OsetReview}. At the lowest order, the isospin
$I=0$ kernel of the
 Bethe-Salpeter equation is
 \begin{eqnarray}
V_{\pi\pi,\pi\pi}^{I=0}(s)&=&-\frac{s-\frac{m_\pi^2}{2}}{f_\pi^2}\,,
 \end{eqnarray}
   and the $D$-wave final-state interactions cannot be included.

The scattering amplitudes for the decays $\psi^{\prime}\to
J/\psi\pi^+\pi^-$ and $\eta_c^{\prime}\to \eta_c\pi^+\pi^-$ can be
written  as
\begin{equation}
T=T^0 + T^0_{S}\cdot G(m_{\pi\pi}^2)\cdot
2T^{I=0}_{\pi\pi,\pi\pi}(m_{\pi\pi}^2)\, ,
\end{equation}
where $2T^{I=0}_{\pi\pi,\pi\pi}=\langle \pi^+\pi^- + \pi^-\pi^+
+\pi^0\pi^0|T^{I=0}|\pi^+\pi^-\rangle$, the $T_{S}^0$ are the
$S$-wave components  of the scattering amplitudes  $T^0$, and the
$G(p^2)$ is the two-meson loop propagator,
\begin{eqnarray}
G(p^2)&=&i\int\frac{d^4q}{(2\pi)^4}\frac{1}{q^2-m_{\pi}^2+i \epsilon} \frac{1}{%
(p-q)^2-m_{\pi}^2+i\epsilon}\, ,\nonumber \\
&=&\frac{1}{(4\pi)^2}\left\{
\widetilde{a}(\mu)+\log\frac{m_\pi^2}{\mu^2}+\sigma\log\frac{\sigma+1}{\sigma-1}\right\}\, ,
\end{eqnarray}
where $p^2=m^2_{\pi\pi}$,
$\sigma=\sqrt{1-\frac{4m_{\pi}^2}{m^2_{\pi\pi}}}$,  and
$\mu=m_\rho=770\,\rm{MeV}$. Here we take the dimensional regulation
to regulate the  ultraviolet divergence and introduce the
subtraction constant $\widetilde{a}(\mu)$  as a free parameter. The
full $S$-wave $\pi\pi\to \pi\pi$ scattering amplitude
$T^{I=0}_{\pi\pi,\pi\pi}$ can be taken as the solution of the
on-shell Bethe-Salpeter equation \cite{OsetReview},
\begin{eqnarray}
T^{I=0}_{\pi\pi,\pi\pi}(m^2_{\pi\pi})&=&\frac{V^{I=0}_{\pi\pi,\pi\pi}(m^2_{\pi\pi})}{1-G(m^2_{\pi\pi})V^{I=0}_{\pi\pi,\pi\pi}(m^2_{\pi\pi})}\,,
\end{eqnarray}
where we have neglected the contributions from the $K\bar{K}$
channels considering the values
$M_{\psi^{\prime}}-M_{J/\psi}<2m_{K}$ and
$M_{\eta_c^{\prime}}-M_{\eta_c}<2m_{K}$.

The differential decay width of the transition $\psi^{\prime}\to
J/\psi\pi^+\pi^-$ can be written as
\begin{equation}
\frac{d\Gamma_{\psi^{\prime}\to
J/\psi\pi^+\pi^-}}{dm_{\pi\pi}}=\frac{1}{(2\pi)^38M_{\psi^{\prime}}^2}
\overline{\sum}\sum|T|^2|{\bf p_1^*}| p_{J/\psi}d\cos{\theta}^*_\pi
\, ,
\end{equation}
where
\begin{eqnarray}
p_{J/\psi}&=&\frac{\sqrt{\left[M_{\psi^{\prime}}^2-(M_{J/\psi}+m_{\pi\pi})^2\right]\left[M_{\psi^{\prime}}^2-(M_{J/\psi}-m_{\pi\pi})^2\right]}}{2M_{\psi^{\prime}}}\,,
 \end{eqnarray}
the $\overline{\sum}\sum$ denotes the average over the polarization
vector of the initial state $\psi^{\prime}$ and the sum over the
polarization vector of the final state $J/\psi$. The corresponding
differential decay width of the transition
$\eta_c^{\prime}\to\eta_c\pi^+\pi^-$ can be obtained with a simple
replacement.

\section{Numerical results and discussions}

The  coupling constants $g_1$, $g_2$ and $g_3$ and the subtraction
constant $\widetilde{a}(\mu)$ can be fitted to  the experimental
data on the transition $\psi^{\prime}\to J/\psi\pi^+\pi^-$ from the
BES collaboration \cite{Exp-BES-jpsi}. In
Refs.\cite{Mannel,ZhuangTL,FKGuo2-1,FKGuo2-2},  the coupling
constant $g_3$ associates with the small $m_{\pi}^2$ is neglected.
In this article, we retain the coupling constant $g_3$ and  fit the
parameters to the experimental data in  four cases: $2{\rm CC}$,
$3{\rm CC}$, $2{\rm CC}+{\rm FSI}$ and $3{\rm CC}+{\rm FSI}$,
respectively,  where the 2CC denotes the two coupling constants
$g_1$ and $g_2$, the 3CC denotes the three coupling constants $g_1$,
$g_2$ and $g_3$, the FSI denotes the final-state interactions. In
the chiral limit, the Adler zero condition can be satisfied.  The
numerical results are plotted as the number of events versus the
$\pi\pi$ invariant momentum $m_{\pi\pi}$, see Fig.1. From the
figure, we can see that retaining only  the coupling constants $g_1$
and $g_2$ can lead to rather satisfactory fitting, by adding the
coupling constant $g_3$ and final-state interactions, even better
fittings  can be obtained.

We normalize  the BES data  using the width $\Gamma_{\psi^{\prime}}=
286\, {\rm keV}$ and the branching ratio ${\rm Br}(\psi^{\prime}\to
J/\psi\pi^+\pi^-)=33.6\%$ \cite{PDG}. The numerical values of the
coupling constants are shown in Table 1, where the unit of the
coupling constants $g_1$, $g_2$ and $g_3$ is $\rm{GeV}^{-1}$, the
subtraction constant $\widetilde{a}(\mu)$ is a dimensionless
quantity. From the Table, we can see that the parameters from the
four cases differ from each other remarkably (or significantly), the
coupling constant $g_3$ and the final-state interactions maybe play
an important role, and we should take them into account.

Using the parameters presented in Table 1, we can obtain the decay
width and the differential decay width of the transition
 $\eta_c^{\prime}\to \eta_c \pi^+\pi^-$ in the four cases $2{\rm CC}$,
$3{\rm CC}$, $2{\rm CC}+{\rm FSI}$ and $3{\rm CC}+{\rm FSI}$, which
are shown in Table 1 and Fig.2, respectively. From the Fig.2, we can
see that the line-shapes of the differential decay width of the
transition  $\eta_c^{\prime}\to \eta_c \pi^+\pi^-$  differ from each
other significantly in the four cases, although those parameters
can all give satisfactory descriptions of the $\psi^{\prime}\to
J/\psi \pi^+\pi^-$ differential decay width.

In Ref.\cite{Voloshin-etac}, M. B. Voloshin studies the transitions
$\psi^{\prime}\to {J/\psi}\pi^+\pi^-$ and $\eta_c^{\prime}\to
{\eta_c}\pi^+\pi^-$ in the framework of the multipole expansion in
QCD using the current algebra and the trace anomaly in QCD, and
obtain the ratio,
\begin{eqnarray}
\frac{\Gamma_{\eta_c^{\prime}\to\eta_c\pi\pi}}{\Gamma_{\psi^{\prime}\to
J/\psi\pi\pi}}&=&3.5\pm0.5\,,
\end{eqnarray}
the lower bound is compatible with the upper bound of the present
prediction $2.50^{+0.35}_{-0.35}$ in the case of the 3CC+FSI. In the
cases of the 3CC and 2CC+FSI, the uncertainties of the present
predictions are too large. We can confront the present predictions
with the experimental data at the BESIII and $\bar{{\rm
P}}\rm{ANDA}$ in the future \cite{BESIII,PANDA}, and  obtain
powerful constraints on the chiral breaking effects and the
final-state interactions, and test the heavy quark symmetry.

\begin{table}
\begin{center}
\begin{tabular}{|c|c|c|c|c|c| }\hline\hline
                                      & 2CC                    & 3CC                       & 2CC+FSI                    &  3CC+FSI            \\ \hline
              $g_1$                   & $0.0873\pm0.0008$      & $0.1086\pm0.0039$         & $0.0586\pm0.0088$          &  $0.0468\pm0.0014$  \\ \hline
              $g_2$                   & $-0.0258\pm0.0010$     & $-0.1814\pm0.0247$        & $-0.0231\pm0.0031$         &  $0.0033\pm0.0018$  \\ \hline
              $g_3$                   &                        & $0.5098\pm0.0787$         &                           &  $-0.0794\pm0.0062$  \\ \hline
              $\widetilde{a}$         &                        &                           & $-1.0661\pm0.5755$         &  $ -1.8160\pm0.1752$  \\ \hline
              $\Gamma_{\eta_c\pi\pi}$ & $229.7^{+7.4}_{-7.3}$  & $140.0^{+121.9}_{-74.0}$  & $209.1^{+123.1}_{-102.6}$  &  $240.3^{+34.1}_{-33.3}$  \\ \hline
    $\widehat{\Gamma}_{\eta_c\pi\pi}$ & $2.39^{+0.08}_{-0.08}$ & $1.46^{+1.27}_{-0.77}$    & $2.18^{+1.28}_{-1.07}$     &  $2.50^{+0.35}_{-0.35}$  \\ \hline   \hline
\end{tabular}
\end{center}
\caption{ The  parameters fitted to the experimental data on the
 $\psi^{\prime} \to J/\psi \pi^+\pi^-$ decay, the unit of the
  $\eta_c^{\prime} \to\eta_c\pi^+\pi^-$ decay width is
KeV, and
$\widehat{\Gamma}_{\eta_c\pi\pi}=\frac{\Gamma_{\eta_c^{\prime}\to\eta_c\pi\pi}}{\Gamma_{\psi^{\prime}\to
J/\psi\pi\pi}}$. }
\end{table}

\begin{figure}
 \centering
 \includegraphics[totalheight=8cm,width=7cm]{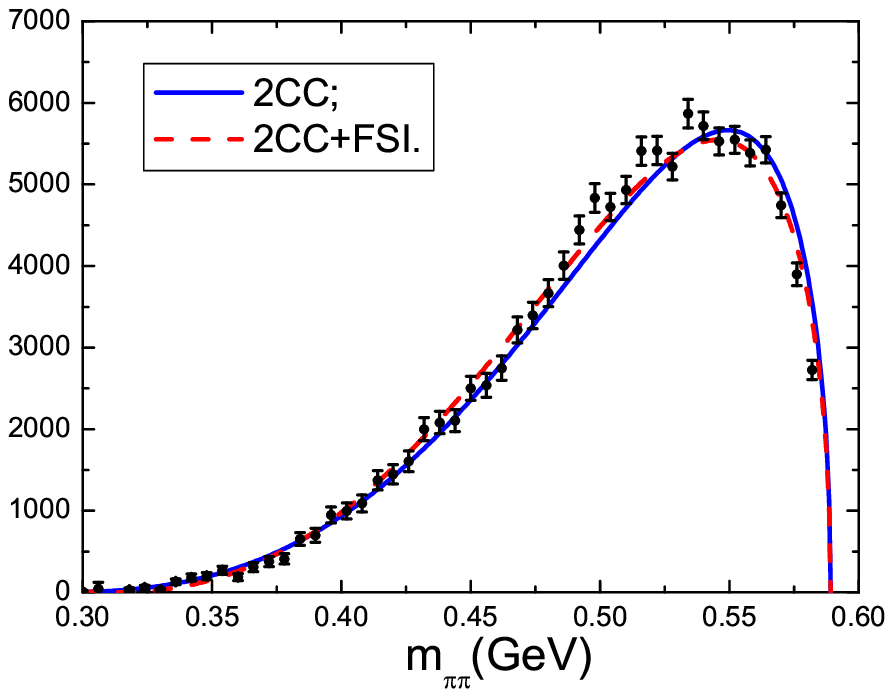}
 \includegraphics[totalheight=8cm,width=7cm]{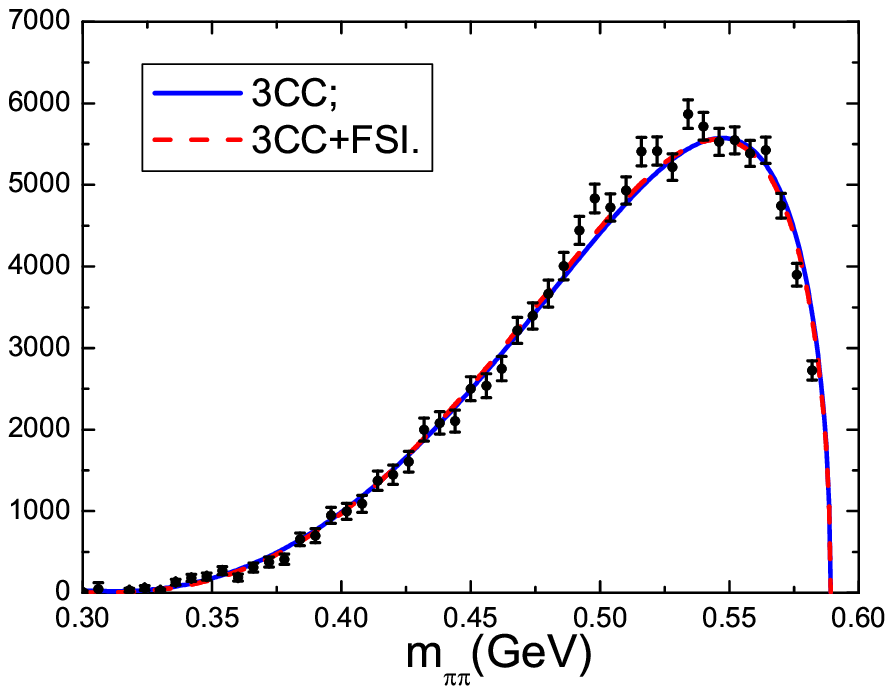}
  \caption{The number of events versus the $\pi\pi$ invariant mass
  distribution   $m_{\pi\pi}$ in the decay $\psi^{\prime}\to J/\psi \pi^+\pi^-$,  the normalizition terms are not shown
  explicitly. The experimental data is taken from the BES collaboration. }
\end{figure}

\begin{figure}
 \centering
 \includegraphics[totalheight=8cm,width=12cm]{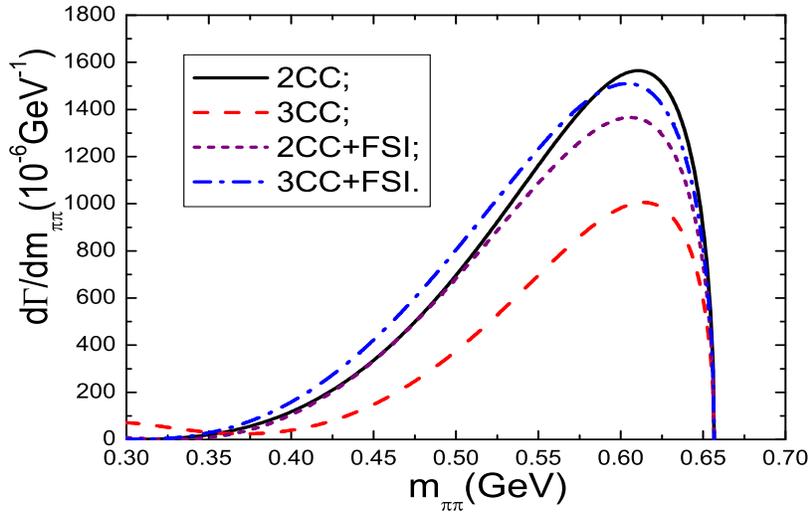}
  \caption{The  differential decay width of the transition $\eta_c^{\prime}\to \eta_c \pi^+\pi^-$ versus the $\pi\pi$ invariant mass
  distribution   $m_{\pi\pi}$.  }
\end{figure}

\section{Conclusion}
In this article, we study the decays $\psi^{\prime}\to
{J/\psi}\pi^+\pi^-$ and $\eta_c^{\prime}\to {\eta_c}\pi^+\pi^-$ by
taking into account the chiral breaking effects,  the final-state
interactions, and the heavy quark symmetry. We fit the parameters to
the experimental data on the $\psi^{\prime}\to {J/\psi}\pi^+\pi^-$
from the BES collaboration, and then take those values to calculate
the decay width and differential decay width of the transition
$\eta_c^{\prime}\to \eta_c \pi^+\pi^-$, which can be confronted with
the experimental data in the future, and put powerful constraints on
the chiral breaking effects and the final-state interactions, and
test the heavy quark symmetry.

\section*{Acknowledgments}
This  work is supported by National Natural Science Foundation,
Grant Number 11075053, and Program for New Century Excellent Talents
in University, Grant Number NCET-07-0282, and the Fundamental
Research Funds for the Central Universities.

\end{document}